\documentclass[a4paper,12pt]{article}

\usepackage{natbib}
\usepackage[utf8]{inputenc}
\usepackage{amsmath}
\usepackage{amsfonts} 
\usepackage{amssymb}
\usepackage{graphicx}

\usepackage[figuresright]{rotating}
\usepackage{multirow}

\usepackage{hyperref}

\begin{document}
\title{Periodic orbits of planets in binary systems}
\author{George Voyatzis \\ Department of Physics, Aristotle University of Thessaloniki, Greece, \\ voyatzis@auth.gr} 

\maketitle

\begin{abstract}
Periodic solutions of the three body problem are very important for understanding its dynamics either in a theoretical framework or in various applications in celestial mechanics. In this paper we discuss the computation and continuation of periodic orbits for planetary systems. The study is restricted to coplanar motion. Staring from known results of two-planet systems around single stars, we perform continuation of solutions with respect to the mass and approach periodic orbits of single planets in two-star systems. Also, families of periodic solutions can be computed for fixed masses of the primaries. When they are linearly stable, we can conclude about the existence of phase space domains of long-term orbital stability.   \\  
{\bf keywords :} periodic orbits, orbital stability, planetary systems, binary systems
\end{abstract}

\section{Introduction}
The study of orbital stability of small celestial bodies in single or multiple star systems is a very important issue in order to understand the dynamical evolution and the origin of our Solar or extrasolar systems. A basic model for such studies is the three body problem (TBP) \citep[e.g., see the review paper of][]{Musielak14} 
but, for numerical studies, $N$-body integrations can be performed, as well.  

For multi-planet systems around single stars, orbital stability can be studied by computing a) Hill's like stability criteria \citep[e.g.,][]{barnes06,veras13a}, b) chaoticity indices and stability maps \citep[e.g.,][]{dvorak03, Erdi04, ckvh07}. c) equilibria in averaged models \citep{mbf06} or periodic orbits \citep{hadjidem84, hadjidem06}. However these methods may be extended also to study the stability of planetary orbits in binary star systems. E.g. in \cite{Szenko08} an improved Hill stability criterion is provided. Extensive numerical simulations and stability limits are given in \citep{Dvorak02, Musielak05}. Periodic orbits in binary systems have studied in the framework of the restricted circular three body problem (CRTBP) in a large number of papers \citep[see e.g.,][and references therein]{Bruno94, Henon97, Broucke01, Nagel08}. However, only few computations of periodic orbits have been performed for elliptic binaries \citep{Broucke69,haghi03}.

Linearly stable periodic orbits consist of centers of foliation of invariant tori in phase space \citep{Berry1978}. Except in cases of circular orbits, they are associated with stable modes of resonances i.e. they are centers of libration of resonant angles and, therefore, indicate dynamical regions of long-term stability. Also, it has been shown that families of periodic orbits are paths of planetary migration when planets migrate due to the interactions with the protoplanetary disk \citep[see e.g,][]{bmfm06, vat14}.  

In the present study we approach the dynamics of planets around binary systems through the computation of periodic orbits. First we review the main aspects (theoretical and computational) of periodic orbits of the planar three body problem and we discuss the continuation of periodic solutions with respect to the mass and location in phase space. Then, we describe how to approach the families of periodic orbits in binary systems by continuing known solutions of the unperturbed problem or of the circular restricted three body problem (CRTBP).

\section{Model and periodic orbits}
We consider the general planar three body problem (GTBP) consisting of three point masses $m_0$, $m_1$ and $m_2$ (bodies $P_0$, $P_1$, $P_2$, respectively), which move under their mutual gravitational interactions on the inertial plane O$XY$, where O is the center of mass. We will assume that $P_0$ is the heaviest body (a star), while each one of the other two bodies may correspond to a planet or a second star in case of a binary system. 
\begin{figure}
	\centering
	\includegraphics[width=0.7\hsize]{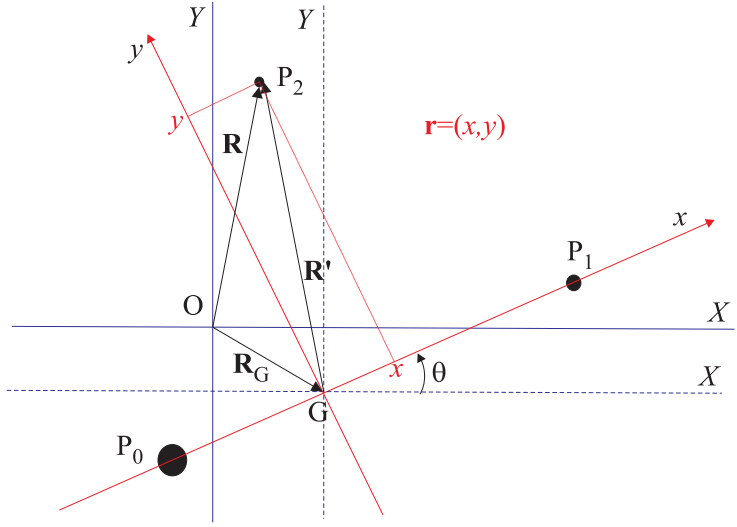}
	\caption{The inertial frame of reference O$XY$, the moving frame G$XY$ (dashed axes) and the rotating frame G$xy$ (red axes).} 
	\label{FigREFFRAMES}
\end{figure}

\subsection{The GTBP in the rotating frame of reference} \label{sectionRModel}
Following \citet{Hadjidem75}, we introduce a rotating frame of reference G$xy$, where $G$ is the center of mass of $P_0$ and $P_1$, the axis $Gx$ is defined by the direction $P_0-P_1$ and the axis G$y$ is vertical to G$x$ (see Fig. \ref{FigREFFRAMES}). The position of the system is given by the coordinates $x_1$ (for $P_1$),  $x_2$, $y_2$ (for $P_2$) and the angle $\theta$ of the rotating axis G$x$ with respect to the inertial one O$X$. For convenience, instead of the variable $x_1$ we will use the distance $r$ between the bodies $P_0$ and $P_1$, 
\begin{equation} \label{RX1}
r=\frac{1-\mu}{x_1}\;>0,\quad \mu=\frac{m_1}{m_0+m_1},
\end{equation}
and the notation $x=x_2$ and $y=y_2$.

If $\mathbf{R}=(X,Y)$ the position vector in the inertial frame O$XY$,  and $\mathbf{R'}=(X',Y')$ the position vector in the frame G$XY$ we have
\begin{equation}\label{TRANSLATION}
\mathbf{R'}=\mathbf{R}-\mathbf{R_G}, \quad \textnormal{or}\quad \mathbf{R'}={\cal T}(\mathbf{R_G})\mathbf{R},   
 \end{equation}
where $\mathbf{R_G}=(m_0\mathbf{R_0}+m_1\mathbf{R_1})/(m_0+m_1)$ and ${\cal T}$ symbolize the translation operator. 
The vector position $\mathbf{r}=(x,y)$ in the rotating frame G$xy$ is given by the rotation of the G$XY$ frame
\begin{equation}\label{ROTATION}
\mathbf{r}={\cal R}(\theta)\mathbf{R'}, \quad\quad {\cal R}(\theta)=\left ( \begin{array}{cc} \cos\theta & \sin\theta \\ -\sin\theta & \cos\theta \end{array} \right ),
\end{equation}
where
\begin{equation}\label{THETA}
\theta = \arctan \left ( \frac{Y_1-Y_0}{X_1-X_0} \right )
\end{equation}

So the transformation of position and velocity from the inertial to the rotating frame is given by the equations
\begin{equation} \label{ROTfromINRT}
\begin{array}{lll}
\mathbf{r}& = &{\cal R}(\theta)\;{\cal T}(\mathbf{R_G})\mathbf{R},\\
\mathbf{\dot r}& = & {\cal R}(\theta)\;{\cal T}(\mathbf{\dot R_G)}\mathbf{\dot R}+\dot \theta {\cal R}'(\theta) {\cal T}(\mathbf{R_G})\mathbf{R},
\end{array}
\end{equation}
where ${\cal R}'(\theta)=\partial R(\theta)/\partial\theta$. The inverse transformation is given by
\begin{equation}\label{INRTfromROT}
\begin{array}{lll}
\mathbf{R} & = &{\cal T}(\mathbf{-R_G})R(-\theta) \mathbf{r}, \\
\mathbf{\dot R}& = &{\cal T}(\mathbf{-\dot R_G})R(-\theta) \mathbf{\dot r}+\dot \theta {\cal T}(\mathbf{-\dot R_G}) [R'(\theta)]^{T} \mathbf{r},
\end{array}
\end{equation}
where $[.]^T$ indicates the transpose matrix.

By applying the transformation (\ref{ROTfromINRT}), and by taking the position and velocity of $P_0$ from the fixed center of mass, the total kinetic energy takes the following form in the rotating frame G$xy$ :
\begin{equation} \label{KE}
{\cal K}=\frac{1}{2} M_1 \left (\dot r^2 +r^2 \dot\theta^2 \right ) + \frac{1}{2} M_2 \left (\dot x^2+\dot y^2 + 2\dot \theta (x \dot y -\dot x y) + \dot\theta^2(x^2+y^2)
\right ),
\end{equation}
where 
$$
M_1=\frac{m_1m_0}{m_1+m_0},\quad M_2=\frac{(m_1+m_0)m_2}{m_0+m_1+m_2},
$$
are the {\em reduced} masses of the system.      

The potential function, with gravitational constant $G$, is
\begin{equation}\label{PE}
{\cal V}=-\frac{G m_0 m_1}{r_{01}}-\frac{G m_1 m_2}{r_{12}}-\frac{G m_0 m_2}{r_{02}},
\end{equation}
where $r_{ij}$ are the distances between the bodies $P_i$ and $P_j$, which are invariant under the tranformation (\ref{ROTfromINRT}) and are written as
$$
r_{01}=r,\quad
r_{12}=\sqrt{\left ( (1-\mu)r-x \right )^2 +y^2},\quad
r_{02}=\sqrt{\left (\mu r+x \right )^2 +y^2}
$$
The equations of motion can be derived from the {\em Lagrangian} function
$$
{\cal L}={\cal K} - {\cal V},
$$ 
we find
\begin{equation} \label{ODEs}
\begin{array}{lll}
\ddot r & =& r \dot\theta^2-\dfrac{G(m_0+m_1)}{r^2}-G m_2 \left [ \dfrac{(1-\mu) r-x}{r_{12}^3}+ \dfrac{\mu r+x}{r_{02}^3} \right ]\\
\ddot x & =& 2\dot\theta \dot y+\dot \theta^2 x +\ddot \theta y+G m \left [\mu\dfrac{(1-\mu)r-x}{r_{12}^3}-(1-\mu)\dfrac{\mu r+x}{r_{02}^3} \right ]\\
\ddot y & =& -2\dot\theta \dot x+\dot \theta^2 y -\ddot \theta x-G m \left [\mu\dfrac{y}{r_{12}^3}+(1-\mu)\dfrac{y}{r_{02}^3}\right ],
\end{array}
\end{equation}
where $m=m_0+m_1+m_2$.    

$\theta$ is an ignorable (cyclic) variable and, therefore, the angular momentum $L=\partial {\cal L}/\partial\dot\theta$ is an integral of motion:
\begin{equation}\label{LANG}
L=M_1 r^2 \dot \theta + M_2 \left (\dot \theta (x^2+y^2) + x\dot y -\dot x y \right )
\end{equation}
If we solve the above equations with respect to $\dot \theta$,  
\begin{equation} \label{THETADOT}
\dot \theta = \frac{L-M_2(x\dot y-\dot x y)}{M_1 r^2 +M_2(x^2+y^2)},
\end{equation}
and substitute in Lagrangian ${\cal L}$, then the position of the system in the rotating frame is defined explicitly by fixing the constant angular momentum $L$ and the system is reduced to three degrees of freedom ($r$, $x$, $y$). However, in order to apply transformation (\ref{INRTfromROT}), we should also know $\theta=\theta(t)$. If we differentiate (\ref{LANG}), substitute the term $x\ddot y - \ddot x y$, which appears in the expressions and can be constructed  by using (\ref{ODEs}), and solving with respect to $\ddot \theta$ we get
\begin{equation} \label{THETADOTDOT}
\ddot \theta = -\frac{2 \dot r \dot\theta}{r}+G m_2\frac{y}{r} \left ( \frac{1}{r_{12}^3}-\frac{1}{r_{02}^3} \right ).
\end{equation}  
So we can integrate the equations (\ref{ODEs}) by integrating simultaneously equation (\ref{THETADOTDOT}). Computationally, it is more convenient to solve the equations of motion in the inertial frame and use transformation (\ref{ROTfromINRT}) to obtain the solution in the rotating frame.  We remark that the origin G of the rotating reference frame does not move uniformly in general.

Apart from angular momentum, the system obeys the Jacobi or energy integral, 
\begin{equation}
{\cal E}={\cal K} + {\cal V}.
\end{equation} 

If we change the scaling of the units such that
\begin{equation}\label{SCALING}
\frac{[m] [x]^3}{[t]^2}=\textnormal{constant}
\end{equation}
the equations (\ref{ODEs}) remain invariant \citep{marchal90}. However, the value of the angular momentum changes correspondingly. In the following, we will consider the normalized mass values 
$$
m_1+m_2+m_3=1, \quad G=1.
$$
Also, the system of equations (\ref{ODEs}) obeys the fundamental symmetry
\begin{equation} \label{EqSymmetry}
\Sigma: (t,r,x,y,\dot r, \dot x, \dot y) \rightarrow (-t,r,x,-y,-\dot r, -\dot x, \dot y).
\end{equation}

By taking the limit $m_2\rightarrow 0$ in the corresponding equations of motion, we obtain the equations of the elliptic restricted three body problem (ERTBP). Furthermore, by considering the consistent solution $r=$const., $\dot\theta=$const., the equations reduce to the equations
of the circular restricted problem (CRTBP).

\subsection{Periodic orbits and stability} \label{SecPOS}
Let $\mathbf{X}=(r,x,y,\dot r,\dot x, \dot y)$ is the position vector in phase space of system (\ref{ODEs}) and $\mathbf{X}=\mathbf{X}(t;\mathbf{X_0})$ defines an orbit with initial conditions $\mathbf{X_0}$=$(r_{0},x_{0},y_{0},\dot r_{0},\dot x_{0}, \dot y_0)$, i.e. a solution of the ODEs (\ref{ODEs}), which are written briefly as
\begin{equation}\label{ODESGEN}
\dot X_i= F_i(X_j),\quad  (X_1,X_2,X_3,X_4,X_5,X_6)\equiv (r,x,y,\dot r,\dot x, \dot y).
\end{equation}
By definition an orbit $\mathbf{X}(t;\mathbf{X_0})$ is periodic of period $T$ if $\mathbf{X}(T;\mathbf{X_0})$=$\mathbf{X_0}$. 

\begin{figure}
	\centering
	\includegraphics[width=0.9\hsize]{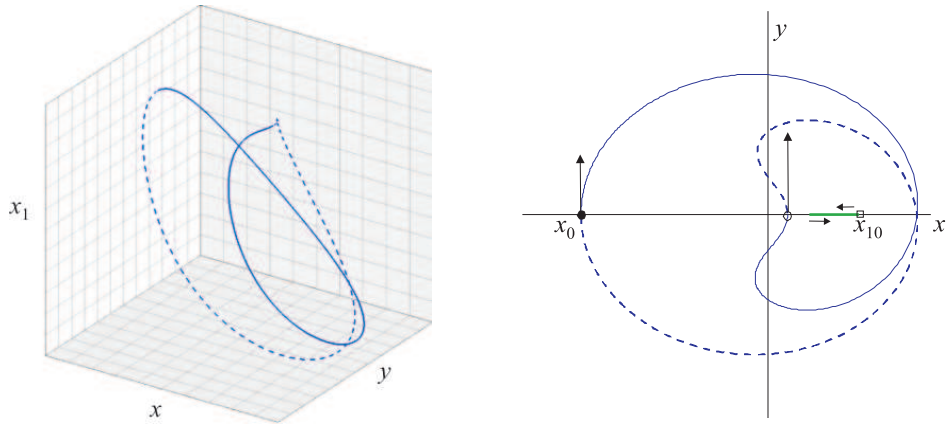}
	\caption{Symmetric periodic orbit in the rotating frame. In the time interval ($T/2, T$) the orbit is presented by the dashed line.  a) presentation in the space ($x_1,x,y$), b) projection of the orbit in $x-y$ plane where the motion of $P_1$ on the $x$-axis is presented by the indicated interval (green). } 
	\label{FigSamplePO}
\end{figure}

A periodic orbit is {\em symmetric} when it is invariant under the symmetry (\ref{EqSymmetry}). I.e. we can always define a symmetric periodic orbit (see Fig. \ref{FigSamplePO}) with initial conditions 
$$y_{0}=\dot r_{0}=\dot x_{0}=0$$
and by assuming the periodicity conditions
\begin{equation}
y(T/2)=\dot r(T/2)=\dot x(T/2)=0.
\end{equation}
Thus, symmetric periodic orbits can be represented by points in the space 
$$\Pi_3=\{(r_{0},x_{0},\dot y_{0})\}.$$  
In the following we will refer only to symmetric periodic orbits. Asymmetric periodic orbits are studied in \citet{vh05} and \citet{avk11}. 
\begin{figure}
	\centering
	\includegraphics[width=0.9\hsize]{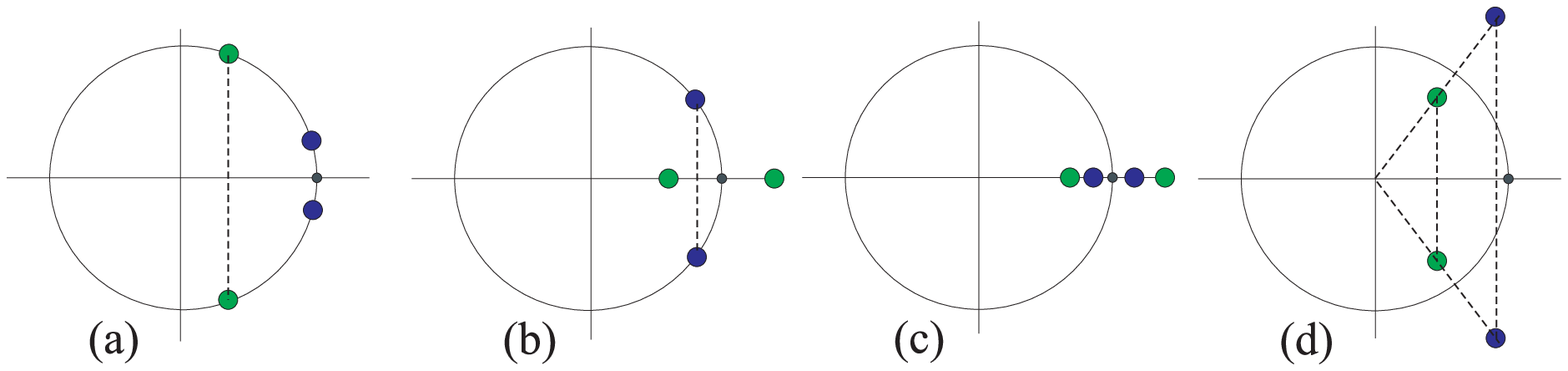}
	\caption{Possible distribution of eigenvalues $\lambda_i$, $i=3,4,5,6$ on the complex plane  a) stability (s) b) single instability (u) c) double instability (uu) d) complex instability (cu). For all cases $\lambda_1$=$\lambda_2=1$} 
	\label{FigEigenvals}
\end{figure}
   
The deviations $\delta X_i$ of the variables $X_i$ along a periodic orbit are given by the {\em variational equations}
\begin{equation}\label{VAREQGEN}
\delta \dot X_i=\sum_{j=1}^6 \left (\frac{\partial F_i}{\partial X_j} \right )_0 \delta X_j,
\end{equation}
where the subscript 0 indicates that the derivatives are computed along the periodic solutions and, therefore, they are periodic functions of $t$. 
If $\mathbf{\Delta}(t)$ is a fundamental matrix of solutions of linear system (\ref{VAREQGEN}), which is called {\em matrizant} or {\em state transition matrix}, the evolution of deviations is given by
$$
\delta \mathbf{X}(t)=\mathbf{\Delta}(t) \delta \mathbf{X}(0),\quad \left (\mathbf{\Delta}(0)=\mathbf{I}_6 \right ).
$$
The evolution of the deviations $\delta X_i(t)$ (bounded or unbounded) depends on the eigenvalues $\lambda_i$, $i=1,..,6$ of the constant matrix 
$\mathbf{\Delta}(T)$, which is called {\em monodromy matrix}. 
Therefore, $\lambda_i$ should define the linear stability of the periodic orbit. Since the planar GTBP is Hamiltonian of 3 d.o.f, the system (\ref{VAREQGEN}) is symplectic and, subsequently, we have always $\lambda_1$=$\lambda_2$=1. The remaining four eigenvalues form reciprocal pairs. Their distribution on the complex plane is presented in Fig. \ref{FigEigenvals} and determines the stability type of the periodic orbit. Actually, only when all eigenvalues  are lying on the unit circle and are different from $-1$ or $1$ (except  $\lambda_1$,$\lambda_2$) the periodic orbit is linearly stable. We remark that linear stability indicates {\em orbital} but not {\em Lyapunov} stability \citep{hadjbook06}.

\section{Continuation of periodic solutions: theoretical aspects}

\subsection{Periodic orbits in the unperturbed system}

\begin{figure}
	\centering
	\includegraphics[width=0.9\hsize]{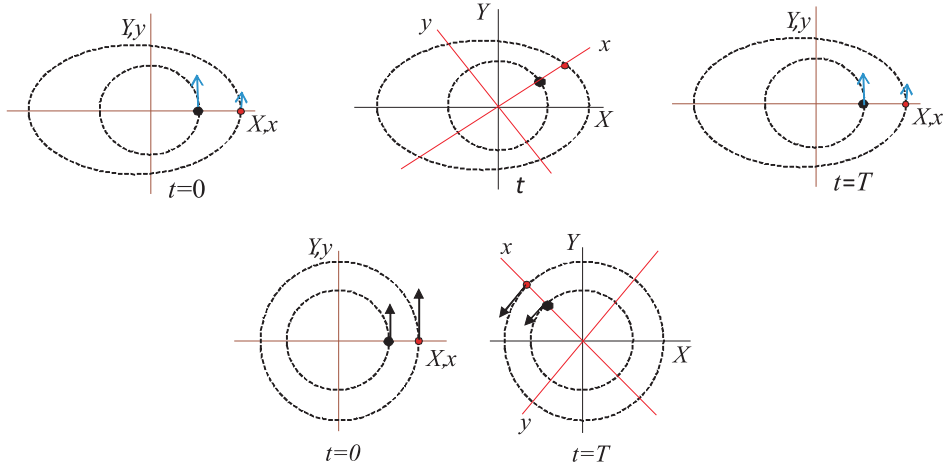}
	\caption{Symmetric periodic planetary configurations for elliptic (above) and circular orbits (below).} 
	\label{FigPConfig}
\end{figure}

Periodic solutions are easily determined for the unperturbed system where $m_0=1$ and $m_1=m_2=0$. Then the bodies $P_1$ and $P_2$ evolve around $P_0$ in Keplerian ellipses with eccentricity $e_i$, semimajor axis $a_i$, period $T_i=2\pi a_i^{3/2}$ and apsidal difference angle $\Delta \varpi=\varpi_2-\varpi_1$, where the index $i=1,2$ stands for the bodies $P_1$ and $P_2$, respectively. If $e_1\neq 0$ or/and $e_2\neq 0$, then we obtain after finite time $t=T$ the same configuration as the initial one only if the orbits are resonant, i.e. the ratio of periods is rational, $T_1/T_2=p/q$, where $p,q$ are co-prime integers (Fig. \ref{FigPConfig})\footnote{We note that if we define the rotating frame by using the body $P_2$, the resonance $p/q$ becomes $q/p$. Any statement given in the following that holds for a resonance $p/q$ also holds for the resonance $q/p$. In the text we present resonances with $p\geq q$.}. Thus for each resonance an infinite number of {\em resonant} periodic solutions is defined since $\Delta \varpi$ is chosen arbitrarily but symmetric periodic orbits are those for $\Delta \varpi=0$ or $\pi$ (aligned or antialigned configuration, respectively). We remark that in this case we have {\em elliptic periodic orbits}, which are defined either in the inertial frame or in the rotating one and have period $T=qT_1=pT_2$. Thus, if we assume the normalization $a_1=1$ or, equivalently, $T_1=2\pi$, the period of elliptic periodic orbits is
$$
T=2 q\pi,  \quad q\in N.
$$                
In case of circular orbits ($e_1=e_2=0$), similarly to the elliptic orbits, we get  periodicity in the inertial frame only when $T_1/T_2=p/q$  or $a_2=(q/p)^{2/3}$. However, in the rotating frame the bodies come periodically to the same relative configuration as the initial one (see Fig. \ref{FigPConfig}) for period
\begin{equation}\label{C0FamT}
T=\frac{2\pi}{n_2-n_1}, \quad (n_1=1,\;\;n_2=a_2^{-3/2}),     
\end{equation}
where $n_i$ indicates the mean motion. Thus, by taking into account the relation  $\dot y=\dot Y_2-r_0 \dot \theta$, the initial conditions 
\begin{equation}\label{C0Fam}
C_0=\{(r_{0},x_{0},\dot y_{0})\}=\{(1,a_2,a_2^{-1/2}-a_2)|\;a_2\in R\}\;\in\;\Pi_3,
\end{equation}
form a monoparametric characteristic curve in $\Pi_3$, which is the {\em circular family} of the unperturbed system. 

\subsection{From the unperturbed system to CRTBP}
By setting $m_1=\mu$ ($0<\mu\ll m_0$, $m_2=0$) and $r=1$, $\dot \theta=1$ we get the CRTBP, which can be assumed as a perturbed Hamiltonian system with perturbation parameter $\mu$. Now, initial conditions of a periodic orbit can be represented by a point in the plane 
$$\Pi_2=\{(x_{0},\dot y_{0})\} \quad \textnormal{or} \quad \Pi_2=\{(x_{0},C_J)\},$$
where $C_J$ is the value of the Jacobi constant. The following theorems for the existence of periodic solutions in the perturbed system hold :
\begin{itemize}
\item All circular periodic solutions of the unperturbed system continue to exist under the small perturbation with period and initial conditions close to the ones of the unperturbed system except those with period $T=2\pi\frac{k+1}{k}$, $k\in N$. Thus the circular family $C_0$ continues to exist in the CRTBP as a family $C$. As $\mu\rightarrow 0$ the period of orbits along $C$ approximates Eq. (\ref{C0FamT}) but the family shows gaps at the resonances of the form $\frac{k+1}{k}$, \citep[see][]{Henon97, hadjbook06}. Also, the circular periodic orbits are linearly stable except those which are close to the resonances of the form $\frac{2k+1}{2k-1}$.
\begin{figure}
	\centering
	\includegraphics[width=0.9\hsize]{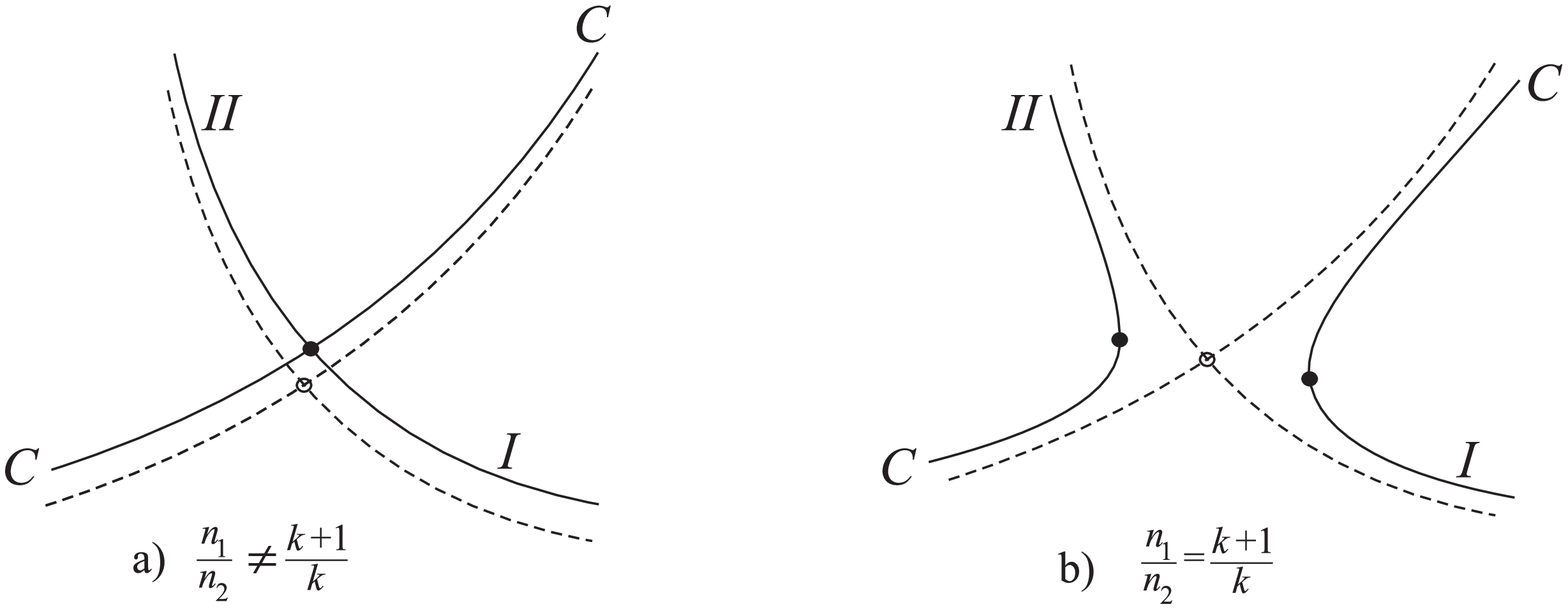}
	\caption{Families of periodic orbits in $\Pi_2$ plane : Continuation of families from the unperturbed (dashed lines) to the perturbed system (solid lines). a) Near a resonance $\frac{n_1}{n_2}\neq\frac{k+1}{k}$. The families of elliptic orbits $I$ and $II$ bifurcate from the resonant orbit on the circular family $C$.  b) Near a resonance $\frac{n_1}{n_2}=\frac{k+1}{k}$. For $\mu\neq 0$ the circular family $C$ breaks and joins smoothly the families of elliptic orbits, $I$ and $II$.} 
	\label{FigPBif}
\end{figure}

\item From the elliptic periodic solutions ($e_2\neq 0$), which exist for any $\Delta\omega\in [0,2\pi)$ in a particular resonance of the unperturbed system, only an even number of them continue to exist under the perturbation. This is concluded from the Poincar\'e-Birkhoff fixed point theorem. Numerical computations show that only two periodic orbits continue to exist for $\mu\neq 0$, particularly those for $\Delta \varpi=0$ and $\Delta \varpi=\pi$  (i.e. the symmetric orbits), and one is stable and the other unstable. By considering $e_2$ as a parameter we obtain two monoparametric families of elliptic orbits, one for $\Delta \varpi=0$ and one for $\Delta \varpi=\pi$, called {\em family I} and {\em family II}, respectively. 
\end{itemize}

The families $C$, $I$ and $II$ are represented by {\em characteristic curves} in the plane $\Pi_2$. We note that the eccentricity $e_2$ refers to the osculating eccentricity for the initial conditions of the periodic orbit and depends on $x_0$ and $\dot y_0$. Along the circular family $C$ the period $T$ varies, but along the elliptic families $I$ and $II$ it remains almost constant close to an integer multiple of $2\pi$ (i.e. the period of the primaries). Thus, elliptic periodic orbits are $p/q$ resonant periodic orbits.  

As $e_2\rightarrow 0$ the elliptic families, $I$ and $II$, meet the circular family $C$ at a resonant circular periodic orbit which is a bifurcation point for the elliptic families. However, for the resonances of the form $\frac{k+1}{k}$, where there exists a gap, the elliptic family joins smoothly the circular family (see Fig. \ref{FigPBif}).

\subsection{From the CRTBP to the ERTBP} {\label{CRTBPtoERTBP}
 By setting the primaries of the RTBP to move on an elliptic orbit ($e_1>0$, $a_1=1$) with period $T_1=2\pi$, we obtain the ERTBP. The ERTBP is structurally different from CRTBP because is non autonomous and the Jacobi integral does not exist. Since the system is periodic in time with period $2\pi$, all its periodic orbits must have period $T=2k\pi$, $k\in N$. The following theorem holds \citep{Broucke69} 
\begin{itemize}
\item All periodic solutions of the CRTBP ($\mu\neq 0$, $e_1=0$) with period $T=2\pi \frac{p}{q}$ continue to exist for $0<e_1\ll1$ with slightly different initial conditions and period $T=2\pi p$. Computations show that continuation may be possible up to $e_1=1$. 
\end{itemize}
By varying $e_1$, the periodic orbits form families $E$, which are represented by characteristic curves in the space of initial conditions
$$\Pi'_3=\{(e_1,x_{0},\dot y_{0})\}.$$ 
The space $\Pi_3$ can be also considered since for symmetric orbits it is $r_0=1\pm e_1$. 
The periodic orbits of the CRTBP which have  periods exactly equal to rational multiples of the period of the primaries are {\em isolated orbits} on the families $C$, $I$ and $II$. These orbits are bifurcation points for the families $E$ of the ERTBP. Furthermore, from each bifurcation point two families, $E_p$ and $E_a$, bifurcate by considering the initial position of the body $P_1$ at pericenter ($r_0=1-e_1$) or apocenter ($r_0=1+e_1$), respectively. Many computations of periodic orbits for KBOs by using the CRTBP and continuation to ERTBP are given in \citet{vk05}.
        
\subsection{From RTBP to GTBP}
We consider the restricted three body problem (RTBP), circular or elliptic with $m_1=\mu$ ($0<\mu\ll 1$), and we set $0<m_2\ll 1$ (GTBP of planetary type). The following statements hold
\begin{itemize}
\item All periodic solutions of the CRTBP with period $T\neq 2k\pi$ continue to exist in the GTBP for $0<m_2\ll 1$ with slightly different initial conditions and the same period \citep{Hadjidem75}. These orbits form the family $G_C$ for a particular value $m_2\neq 0$, which is presented by a smooth curve in the space $\Pi_3$. 
\item All periodic orbits of the ERTBP, which belong to a family $E$, are continued in the GTBP for $0<m_2\ll 1$ \citep{HadjiCrist75, avk11}. For a particular value $m_2\neq 0$ they form a family $G_E$, which is also presented by a smooth curve in the space $\Pi_3$ located close to the family $E$.
\end{itemize}
Actually only the periodic orbits of the CRTBP that are bifurcation points for the families $E$ of the ERTBP are not continued. 
At the critical orbits, which are not continued in the GTBP, the families $G_C$ and $G_E$ join smoothly and form two family segments separated by a gap \citep[see Fig. 4 in][]{vkh09}.

\section{Continuation of periodic orbits: computational aspects}
For the computation of periodic orbits we can use the method of differential corrections which is applied through a Newton-Raphson shooting method. Continuation is defined in two ways i) continuation by varying the mass parameter ($\mu$-continuation) ii) continuation in phase space with fixed mass parameter ($x$-continuation). The application of the method requires a starting periodic orbit. Therefore, from a theoretical point of view, we should start the computations from orbits of the unperturbed system and then perform $\mu$-continuation.    

\subsection{Computations in the CRTBP} \label{CompCRTBP}
Let us assume a solution of the CRTBP with mass parameter $\mu_0$ and initial conditions ($x_{0},y_{0},\dot x_{0},\dot y_{0}$) :
\begin{equation}
\begin{array}{ll}
x=x(t;x_{0},y_{0},\dot x_{0},\dot y_{0};\mu_0),\quad y=y(t;x_0,y_0,\dot x_0,\dot y_0;\mu_0),\\
\dot x=\dot x(t;x_0,y_0,\dot x_0,\dot y_0;\mu_0),\quad \dot y=\dot y(t;x_0,y_0,\dot x_0,\dot y_0;\mu_0).
\end{array}
\end{equation}
We consider a symmetric periodic orbit,  
\begin{equation} \label{PODEF1}
\begin{array}{ll}
x(T;x_0,0,0,\dot y_0;\mu_0)=x_0, \quad y(T;x_0,0,0,\dot y_0;\mu_0)=0,\\
\dot x(T;x_0,0,0,\dot y_0;\mu_0)=0, \quad\dot y(T;x_0,y_0,\dot x_0,\dot y_0;\mu_0)=\dot y_0,
\end{array}
\end{equation}
which is determined by the initial conditions ($x_0,\dot y_0$) on a Poincar\'e surface of section $y=0$ and the periodicity condition
\begin{equation} \label{PCONT1}
\dot x(t_s;x_0,0,0,\dot y_0;\mu_0)=0.
\end{equation}
$t_s$ is an appropriate time of a section crossing. In the time interval $(0,T/2]$ the orbit should cross the section $y=0$ for $l\geq 1$ times. This number of crossings defines the {\em multiplicity} of the symmetric periodic orbit.  In computations $l$ must be known (or declared) a priori. Thus $t_s$ is the time at the $l$-th sequent section crossing and, if (\ref{PCONT1}) is satisfied, then $t_s=T/2$. In most cases $l$ is considered to be the multiplicity of the starting periodic orbit but there are cases where continuation takes place with larger multiplicity (and, obviously, with a multiple period).            

\subsubsection{$\mu$-continuation} \label{SectionMuCont1}
Let us assume the periodic solution (\ref{PODEF1}) and search for a new periodic solution for $\mu_1=\mu_0+\delta\mu$. As we have mentioned, for any fixed value of $\mu$, periodic orbits form a continuous set of solutions (a family) in the plane $\Pi_2=\{(x_0,\dot y_0)\}$. Thus, continuation to a unique new periodic orbit have to be defined explicitly e.g. by assuming the same $x_0$ value as initial condition for the new orbit\footnote{We can also seek for a new periodic orbit with the same period.}. Subsequently, we are seeking only for a new initial value for $\dot y$, say $\dot y_0+\delta\dot y_0$, such that the periodicity condition (\ref{PCONT1}) holds,
\begin{equation} \label{PCONT2}
\dot x(t_s;x_0,0,0,\dot y_0+\delta\dot y_0;\mu_1)=0,
\end{equation}
If we consider that $\delta\mu$ is small, we expect that $\delta\dot y_0$ is small too. Thus expanding (\ref{PCONT2}) up to first order with respect to $\delta\dot y_0$ we get
$$
\dot x(t_s;x_0,0,0,\dot y_0;\mu_1)+\left . \frac{\partial \dot x}{\partial \dot y_0} \right |_{t=t_s} \delta \dot y_0 +O(\delta \dot y_0^2)=0
$$
or
\begin{equation}\label{DCORR1}
\delta \dot y_0\approx - \left (\left . \frac{\partial \dot x}{\partial \dot y_0} \right |_{t=t_s} \right )^{-1} u_0, \quad u_0=\dot x(t_s;x_0,0,0,\dot y_0;\mu_1). 
\end{equation}

The above equation gives the first correction for the new periodic orbit, which may not be sufficient due to first order approximation applied. Therefore we repeat the procedure $n$ times, with initial condition $\dot y^{(1)}_{0}=\dot y_0+\delta \dot y_0$, $\dot y^{(2)}_{0}=\dot y^{(1)}_{0}+\delta \dot y_0$ etc, until
\begin{equation} \label{ACCCOND}
 \dot x(t_s;x_0,0,0,\dot y^{(n)}_{20};\mu_1) < \textnormal{tolerance}. 
\end{equation}
Note that all numerical integrations are performed in the interval $t\in [0,t_s]$, where $t_s$ changes when initial conditions of integration change. When (\ref{ACCCOND}) is satisfied, $2t_s$ indicates the best approximation for the period $T$ of the new periodic orbit. 

The whole procedure is repeated for $\mu_2=\mu_1+\delta\mu$ etc. and a $\mu$-{\em family} is constructed at a particular value $x=x_{0}$. We can start the procedure with $m_1=0$ and increasing $m_2$. So, we obtain a $\mu$-family of the CRTB starting from the unperturbed problem, where all periodic solutions are analytically known.

\subsubsection{$x$-continuation} \label{SectionXCont1}
Now we consider a periodic orbit $(x_0,\dot y_0)\in\Pi_2$, for a particular value of $\mu$ i.e. the orbit is a member of a $\mu$-{\em family}. Since a family for fixed $\mu$ is represented by a curve in $\Pi_2$, we can write that $\dot y_0=f(x_0)$, where $f(x)$ is a continuous function and, at least locally, single-valued. Thus, we can compute a new periodic orbit at $x_0+\delta x_0$ by computing the corresponding $\dot y_0+\delta \dot y_0$ using differential corrections. The new periodic orbit must satisfy the periodicity condition
$$
\dot x(t_s;x_0+\delta x_0,0,0,\dot y_0+\delta\dot y_0)=0,    
$$
and by taking $\delta x_0\ll 1$ and assuming that $\delta \dot y_0\ll 1$, too, we write
$$
\dot x(t_s;x_0+\delta x_0,0,0,\dot y_0)+\left . \frac{\partial \dot x}{\partial \dot y_0} \right |_{t=t_s} \delta \dot y_0 +O(\delta \dot y_0^2)=0,    
$$  
and get
\begin{equation} \label{DCORR2}
\delta \dot y_0\approx -\left (\left . \frac{\partial \dot x}{\partial \dot y_0} \right |_{t=t_s} \right )^{-1} u_0, \quad u_0=\dot x(t_s;x_0+\delta x_0,0,0,\dot y_0). 
\end{equation}
Iterating the procedure and provided it is convergent, we obtain the requested corrected initial conditions and the period when
$$
\dot x(t_s;x_0+\delta x_0,0,0,\dot y^{(n)}_{0}) < \textnormal{tolerance}    \quad (T\simeq 2t_s).
$$
The whole procedure is repeated for $x^{(n)}_0=x_0+n\delta x_0$, $n=1,2,3,...$ and a $x$-{\em family}, for fixed $\mu$, is constructed as a characteristic curve in the $\Pi_2$ plane.

\subsubsection{Some technical remarks}	\label{TECHREM}

{\em Computation of derivatives}\\ 
It can be shown \citep{hadjbook06} that the elements of the matrizant $\mathbf{\Delta}(t)=\left (\Delta_{ij}(t)\right )$ are given by  
$$
\Delta_{ij}(t)=\frac{\partial X_i(t)}{\partial X_{j0}},\quad X_{j0}=X_j(0).
$$
Therefore the derivatives which are required for the continuation method (e.g. $\frac{\partial \dot x}{\partial \dot y_0}$ in (\ref{DCORR1}) and (\ref{DCORR2})) can be computed from the solution of variational equations (\ref{VAREQGEN}) at $t=t_s$ and obtaining the deviations $\Delta X_i(t_s)$. Particularly, for the CRTBP we have a system of four equations of motion for the variables $\mathbf{X}=(x,y,\dot x, \dot y)$. By solving the corresponding variational equations for initial conditions $\delta X_j(0)=(0,0,0,1)$, the derivative in (\ref{DCORR1}) and (\ref{DCORR2}) is given by 
$$
\frac{\partial \dot x}{\partial \dot y_0} = \delta X_3(t_s). 
$$	

It may be more convenient (and quite efficient) if we compute the derivatives directly by numerical integration of the system of ODEs, namely
\begin{equation}
\left (\frac{\partial X_i(t)}{\partial X_{j0}}\right )_{t=t_s}\simeq\frac{X_i(t_s;X_{k0},X_{j0}+\epsilon) - X_i(t_s;X_{k0},X_{j0}-\epsilon)}{2\epsilon},\quad k\neq j.
\end{equation}
In computations, generally, we set $\epsilon\approx 10^{-6}$. Of course, more advanced numerical methods for the estimation of derivatives can be used, \citep[see e.g.][]{Press02}. 
\smallskip

\noindent {\em Extrapolation for global family computation}\\
The families of periodic orbits after $\mu$-continuation or $x$-continuation, may not be described globally by single valued functions  $\dot y_0=f(\mu;x_0=\textnormal{const.})$ or $\dot y_0=f(x_0;\mu=\textnormal{const.})$, respectively. Thus maybe the global characteristic curves of the families can't be constructed by monotonically increasing or decreasing the parameter $\mu$ or $x_0$ of the family. We can overcome this problem by assuming as parameter along the family the length $s$ of the characteristic curve from the starting point. Suppose e.g. that we have computed the first points along a family after $x$-continuation by increasing (or decreasing) the parameter $x_0$ and we get the periodic orbits
$$
P_i=(x^{(i)}_{0},\dot y^{(i)}_{0}),\quad i=0,1,2,..,n_0. 
$$
If $ds_k$ ($k>0$) indicates a {\em distance} between the points $P_{k-1}$ and $P_k$ then each periodic point $P_i$ on the family has a distance from $P_0$ equal to $s_i=\sum_{k=1}^{i}ds_k$. $s_i$ monotonically increases along the family and can be used as the parameter of the family such that
\begin{equation}
x^{(i)}_{0}=p_1 (s_i), \quad \dot y^{(i)}_{0}=p_2 (s_i).
\end{equation}
The functions $p_1(s)$ and $p_2(s)$ can be locally defined, at the $i$th periodic orbit, by a polynomial interpolation function of $s$, which is constructed from the $n_0+1$ points $P_{i-n_0}$, ...,$P_{i}$. Then the $P_{i+1}$ periodic orbit, for $s_{i+1}=s_i+\delta s$, is sought near the initial conditions $(x_0,\dot y_0)$=$(p_1(s_{i+1}),p_2(s_{i+1}))$ (extrapolating values). The step $\delta s$ should be sufficiently small for achieving convergence to the periodic orbit.

\subsection{Computations in the GTBP}
 
Let us consider a solution of the GTBP for masses $m_1=m_{10}$ and $m_2=m_{20}$ ($m_0=1-m_1-m_2$) and initial conditions ($r_0,x_{0},y_{0},\dot r_0,\dot x_{0},\dot y_{0}$). 
This solution is a symmetric periodic orbit of period $T$ if 
\begin{equation} \label{GPODEF}
\begin{array}{l}
r(T;r_0,x_0,0,0,0,\dot y_0;m_{10}, m_{20})=r_0,\\
x(T;r_0,x_0,0,0,0,\dot y_0;m_{10}, m_{20})=x_0,\\
\dot y(T;r_0,x_0,0,0,0,\dot y_0;m_{10}, m_{20})=\dot y_0,
\end{array}
\end{equation}
which are satisfied when the following periodicity conditions hold
\begin{equation} \label{GPODEF1}
y(T/2;r_0,x_0,0,0,0,\dot y_0;m_{10}, m_{20})=0,
\end{equation}
and
\begin{equation} \label{GPODEF2}
\begin{array}{l}
\dot r(T/2;r_0,x_0,0,0,0,\dot y_0;m_{10}, m_{20})=0\\
\dot x(T/2;r_0,x_0,0,0,0,\dot y_0;m_{10}, m_{20})=0.
\end{array}
\end{equation}
The periodicity condition (\ref{GPODEF1}) defines a surface of section in the 6-dimensional phase space, and a symmetric periodic solution always crosses this section due to the symmetry (\ref{EqSymmetry}). Thus we can determine initial conditions for a periodic orbit on the surface of section $y=0$ i.e. in the 5-dimensional space 
$$\Pi_{3\times 2}=\{(r,x,\dot y,m_1,m_2)\}.$$
We define the {\em multiplicity} $l$ of the orbit as in the case of the CRTBP (see section \ref{CompCRTBP}) i.e. $l$ is the number of crossings of the periodic orbit with the section $y=0$ in half period. For a predefined value of $l$ we determine (along the numerical integration of the orbit) the time $t_s$ after $l$ intersections of the orbit with the section. For a symmetric periodic solution $t_s=T/2$. We remark that computations in ERTBP are similar with these in GTBP\footnote{In ERTBP the period of orbits $T$ and, consequently, the section cross time $t_s$, is known a priori.}. 
  
\subsubsection{$\mu$-continuation}
Since we are dealing with monoparametric continuation, we fix the value of $m_1$ or $m_2$ and vary the other one. Generally we can define one mass parameter, $\mu$, such that 
$$m_1=f_1(\mu)\quad  \textnormal{and} \quad m_2=f_2(\mu),$$ 
where $f_1$, $f_2$ are monotonic functions.             
 
Let us assume the periodic solution (\ref{GPODEF}), which corresponds to the mass parameter, say $\mu_0$, and search for a new periodic solution for $\mu_1=\mu_0+\delta\mu$. As we have mentioned in section \ref{SecPOS}, for any fixed value of $\mu$ (or, equivalently, fixed $m_1$ and $m_2$), periodic orbits form a continuous set of solutions (a family) in the space $\Pi_3=\{(r_0,x_0,\dot y_0)\}$. In order to obtain a unique new periodic solution for $\mu_1=\mu_0+\delta\mu$ we may assume fixed the initial condition $r(0)=r_0$ and seek for new initial conditions $x_0+\delta x_0$ and $\dot y_0+\delta \dot y_0$ such that the periodicity conditions (\ref{GPODEF2}) are satisfied at the predefined $l$-th intersection of the orbit with the section $y=0$ at $t=t_s$ :
\begin{equation} \label{GPOCONT1}
\begin{array}{ll}
\dot r(t_s;r_0,x_0+\delta x_0,0,0,0,\dot y_0+\delta \dot y_0;\mu_1)= & 0\\
\dot x(t_s;r_0,x_0+\delta x_0,0,0,0,\dot y_0+\delta \dot y_0;\mu_1)= & 0.
\end{array}
\end{equation}
From eq. (\ref{GPOCONT1}) we can determine in first order approximation the corrections $\delta x_0$ and $\delta \dot y_0$ by considering the first order expansions
\begin{equation} \label{GPOEXP}
\begin{array}{l}
\dot r(t_s;r_0,x_0,0,0,0,\dot y_0;\mu_1)+\left (\dfrac{\partial \dot r}{\partial x_0} \right )_0 \delta x_0 + \left (\dfrac{\partial \dot r}{\partial \dot y_0} \right )_0 \delta \dot y_0 = 0,\\
\dot x(t_s;r_0,x_0,0,0,0,\dot y_0;\mu_1)+\left (\dfrac{\partial \dot x}{\partial x_0} \right )_0 \delta x_0 + \left (\dfrac{\partial \dot x}{\partial \dot y_0} \right )_0 \delta \dot y_0 = 0,
\end{array}
\end{equation}
where the subscript '0' indicates that derivatives are computed for the solution with initial conditions $(r_0,x_0,0,0,0,\dot y_0)$, mass parameter $\mu_1$ and at $t=t_s$. Thus, first order corrections are given by 
\begin{equation} \label{GPOCOR}
\left ( \begin{array}{c} \delta x_0 \\ \delta \dot y_0 \end{array} \right ) = -
\left ( \begin{array}{cc} \frac{\partial \dot r}{\partial x_0} & \frac{\partial \dot r}{\partial \dot y_0} \\
\frac{\partial \dot x}{\partial x_0} & \frac{\partial \dot x}{\partial \dot y_0} \end{array} \right )^{-1}_0 \left ( \begin{array}{c} u_0 \\ v_0 \end{array} \right )
\end{equation}
with $u_0=\dot r(t_s;r_0,x_0,0,0,0,\dot y_0;\mu_1)$ and $v_0=\dot x(t_s;r_0,x_0,0,0,0,\dot y_0;\mu_1)$.

Since the corrections $\delta x_0$ and $\delta \dot y_0$ have been computed in first order approximation, we repeat the computation for the corrected initial conditions $x^{(1)}_0=x_0+\delta x_0$ and $\dot y^{(1)}_0=\dot y_0+\delta \dot y_0$ and obtain new corrections $\delta x^{(1)}_0$ and $\delta \dot y^{(1)}_0$. 
The procedure stops after $n$ iterations when
\begin{equation} \label{GPOACC}
|\dot r(t_s;r_0,x^{(n)}_0,0,0,0,\dot y^{(n)}_0;\mu_1)|+|\dot x(t_s;r_0,x^{(n)}_0,0,0,0,\dot y^{(n)}_0;\mu_1)|<\textnormal{tolerance}. 
\end{equation}

By computing the periodic orbit for $\mu_1=\mu_0+\delta\mu$ we apply the same procedure to compute the periodic orbit for $\mu_2=\mu_1+\delta\mu$ etc. and we form a $\mu$-family of periodic orbits, which can be depicted as a characteristic curve in the 3-dimensional space 
$$\Pi_{2\times 1}=\{(x_0,\dot y_0,\mu)\}.$$
Folding of the characteristic curve may exist and, for these cases, the application of polynomial fitting and extrapolation is necessary for continuation (see section \ref{TECHREM}). 

In the above computations, we imply that the angular momentum (\ref{LANG}) is constant. However, by varying the mass parameter, the preservation of the angular momentum may require relatively large corrections and the convergence of the procedure fails after few steps of the mass parameter variation. In order to overcome this descrepancy we consider a new value of the angular momentum at each $n$-step of the $\mu$ increment as
$$
L=L(r_0,x^{(n-1)}_0,0,0,0,\dot y^{(n-1)}_0;\mu_n).
$$   
Also, instead of $L$ we may keep constant the initial angular velocity $\dot\theta_0$ along the family. In general we can always apply appropriately the scaling of units according to (\ref{SCALING}).

\subsubsection{$x$-continuation}
We consider the system with fixed masses and fixed angular momentum. As we mentioned in section \ref{SecPOS}, symmetric periodic orbits form a monoparametric family in phase space, which is depicted as a characteristic curve in the space $\Pi_3=\{(r_0,x_0,\dot y_0)\}$. We may assume $r_0$ as the parameter of the family. By starting from a known periodic orbit $(r_0,x_0,\dot y_0)$ we are seeking for a periodic orbit at a given $r_0^{(1)}=r_0+\delta r_0$. If the new orbit corresponds to initial conditions ($x_0+\delta x_0$, $\dot y_0+\delta\dot y_0$), the periodicity conditions, which should be satisfied, are
\begin{equation}
\begin{array}{l}
\dot r(t_s;r_0+\delta r_0,x_0+\delta x_0,0,0,0,\dot y_0+\delta \dot y_0)=0,\\
\dot x(t_s;r_0+\delta r_0,x_0+\delta x_0,0,0,0,\dot y_0+\delta \dot y_0)=0.
\end{array}
\end{equation}
By expanding the above conditions up to first order around $(x_0,\dot y_0)$ and solving with respect to the corrections $\delta x_0$, $\delta\dot y_0$ we have the solution (\ref{GPOCOR}) with
$$
u_0=\dot r(t_s;r_0+\delta r_0,x_0,0,0,0,\dot y_0), \quad v_0=\dot x(t_s;r_0+\delta r_0,x_0,0,0,0,\dot y_0).
$$
As in the previous cases, the above computation is repeated until a tolerance condition like (\ref{GPOACC}) is satisfied. The family is constructed by computations in successive steps $r_0^{(i+1)}=r_0^{(i)}+\delta r_0$.

\section{Circumbinary periodic orbits for the restricted problem}

\subsection{Families in the CRTBP}
We consider the circular family of the unperturbed system, $C_0$, which is given by (\ref{C0Fam}), and choose a reference orbit, e.g. for $a_2=3.0$($=x_0$). In the rotating frame, $\dot\theta=1$, we get  $\dot y_0 \approx -2.423$. We perform $\mu$-continuation, as it is described in section \ref{SectionMuCont1}, starting from $\mu=0$ and obtain the $\mu-\mathrm{family}$ $C_\mu(x_0=3.0)$. All of its orbits are almost circular, linearly stable and the variation of $\dot y$ and $T$ along the family is very small (see Fig. \ref{FigCmudytper}).     
\begin{figure}
	\centering
	\includegraphics[width=1.0\hsize]{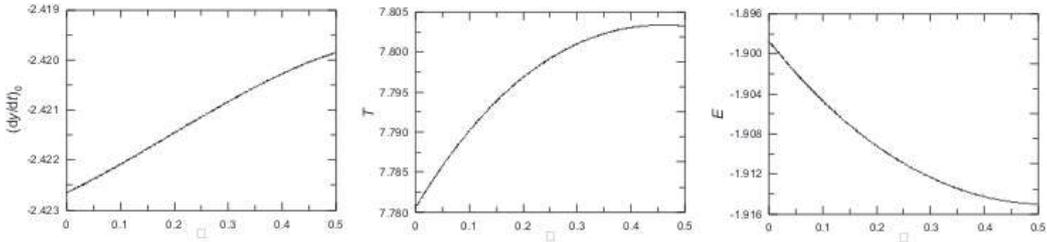}
	\caption{The variation along the family $C_\mu(x_0=3.0)$ of the initial $\dot y_0$, the period $T$ and the energy $E$.}
	\label{FigCmudytper}
\end{figure}

All orbits of the above family $C_\mu$ are continued in phase space with fixed $\mu$. We follow the procedure, which is described in section \ref{SectionXCont1}, and we continue the orbits for $\mu=0.1$, $0.25$ and $0.5$ by varying $x_0$ (starting from $x_0=3.0$). The characteristic curves of the families (which are denoted by $C_x(\mu)$) are presented in Fig. \ref{FigFamCXYdot}. The families start consisting of circular stable orbits but, as $x_0$ decreases, they turn to become unstable and terminate at a very unstable orbit. The termination orbit for $\mu=0.25$ is presented in Fig. \ref{FigTerminationOrbit} in the inertial and in the rotating frame. Although the orbit of the massless body $P_2$ evolves for a long time close to the orbit of $P_1$, actually as it is shown in the rotating frame (where $P_2$ is fixed at $x=0.75$) there is no close encounter. We should notice that for $\mu=0.1$ and $\mu=0.25$, the stable segment of the family (blue) is interrupted by a short segment of unstable orbits (red). 
The period along the families is presented in Fig. \ref{FigCFAMperecc}a. We observe that for about $x_0>1.8$ the period does not depend significantly on the mass parameter $\mu$. As $P_2$ is approaching the binary system, the period starts to increase rapidly and its dependence on $\mu$ becomes clear. In Fig. \ref{FigCFAMperecc}b we present the osculating eccentricity of the periodic orbits, computed at initial conditions with reference to the barycentric system. 

\begin{figure}
	\centering
	\includegraphics[width=1.0\hsize]{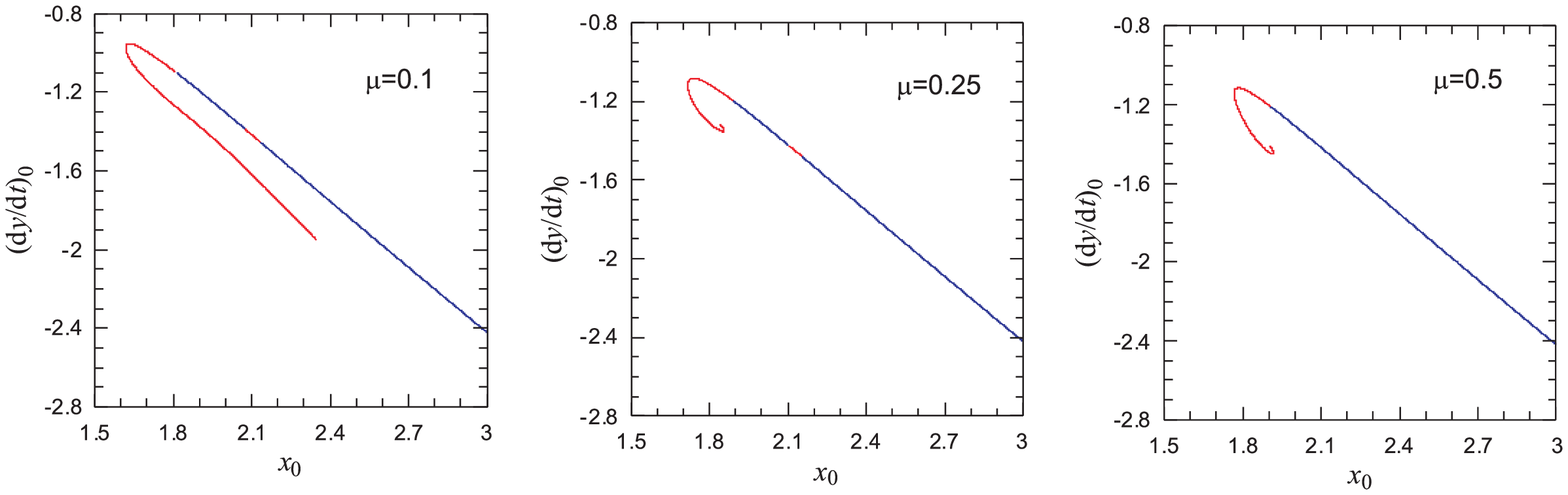}
	\caption{The characteristic curves of the families $C_x(\mu)$ in the plane  $\Pi_2=\{(x_{0},\dot y_{0})\}$  for the indicated values of the mass parameter $\mu$. Blue (red) segments indicate linearly stable (unstable) orbits.} 
	\label{FigFamCXYdot}
\end{figure}
\begin{figure}
	\centering
	\includegraphics[width=0.95\hsize]{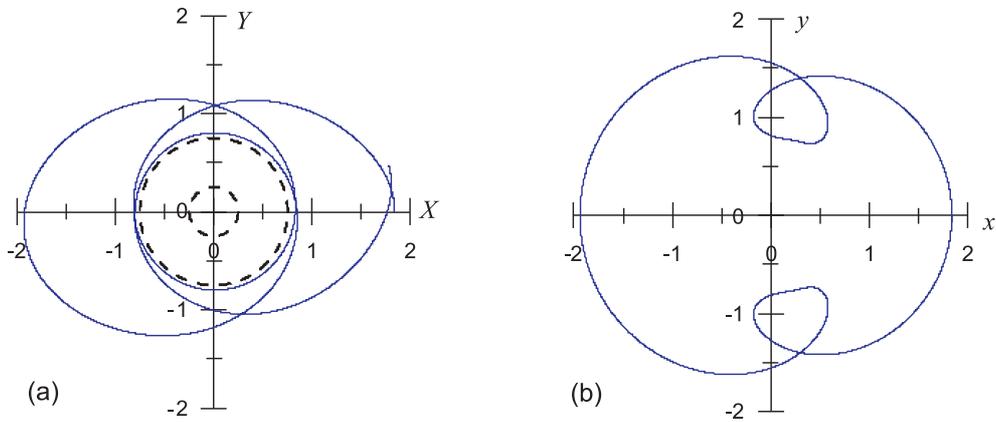}
	\caption{The termination orbit of the family $C_x(\mu=0.25)$ a) in the inertial frame, where dashed circles indicate the orbits of the primaries b) in the rotating frame. We note that the orbit is periodic in the rotating frame but not in the inertial one.} 
	\label{FigTerminationOrbit}
\end{figure}
\begin{figure}
	\centering
	\includegraphics[width=0.95\hsize]{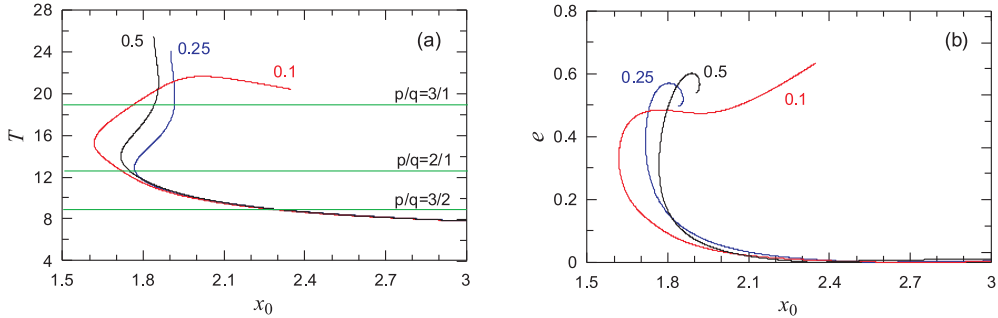}
	\caption{The evolution of   the  period (a) and  the eccentricity (b) along the families $C_x(\mu)$ for $\mu=0.1$, $0.25$ and $0.5$. The lines indicate the values of $T$ where the resonances 3/2, 2/1 and 3/1 exist.} 
	\label{FigCFAMperecc}
\end{figure}
For $x_0>3.0$, the orbits are almost circular. Since the body is moving further away from the binary system as $x_0$ increases, the period $T$ is approximated by equation (\ref{C0FamT}). Thus, it is $T>2\pi$ and $\lim_{x_0\rightarrow \infty}T=2\pi$. For $x_0<2.4$ we observe (Fig. \ref{FigCFAMperecc}b) that the eccentricity starts to increase and, finally, reaches high values. But all these very eccentric orbits belong to the unstable part of the families.

\subsection{Families in the ERTBP}
As we mentioned in section \ref{CRTBPtoERTBP}, the orbits of the families $C_x(\mu)$ with period $T=2\pi\frac{p}{q}$, called {\em generating orbits}, are continued in the elliptic restricted problem ($e_1>0$). From each generating orbit, we obtain the families $E_p$ and $E_a$ with orbits of constant period $T=2p\pi$.  
From Fig. \ref{FigCFAMperecc}a we can observe the existence of generating orbits with $\frac{p}{q}=\frac{3}{2}$, $2$ and $3$. Certainly, an infinite number of generating orbits occurs but in applications we are interested for resonances with small integers $p$ and $q$.

We continue, with respect to the eccentricity of the binary, $e_1$, the generating $3/2$ resonant orbit of the family $C_x(\mu=0.5)$. The bifurcation of the families $E_p$ and $E_a$ from the generating orbit of the circular problem is drawn in the space $\Pi'_3$ and presented in Fig. \ref{FigEfams} (left panel). The eccentricity, $e$, of the circumbinary periodic orbits along the families is presented in the right panel of Fig. \ref{FigEfams}. Family $E_a$ continues up to very high value of eccentricity $e_1$. It starts having single unstable orbits and, after a short segment of double instability, the orbits become complex unstable. The family tends to terminate at a collision orbit.  Family $E_p$ starts, also, with single unstable orbits, which for $e_1>0.49$ become doubly unstable. As $e_1\rightarrow 0.6$ the orbits are very unstable and the convergence of the differential corrections becomes very slow.  
\begin{figure}
	$\begin{array}{cc}
	\includegraphics[width=0.4\hsize]{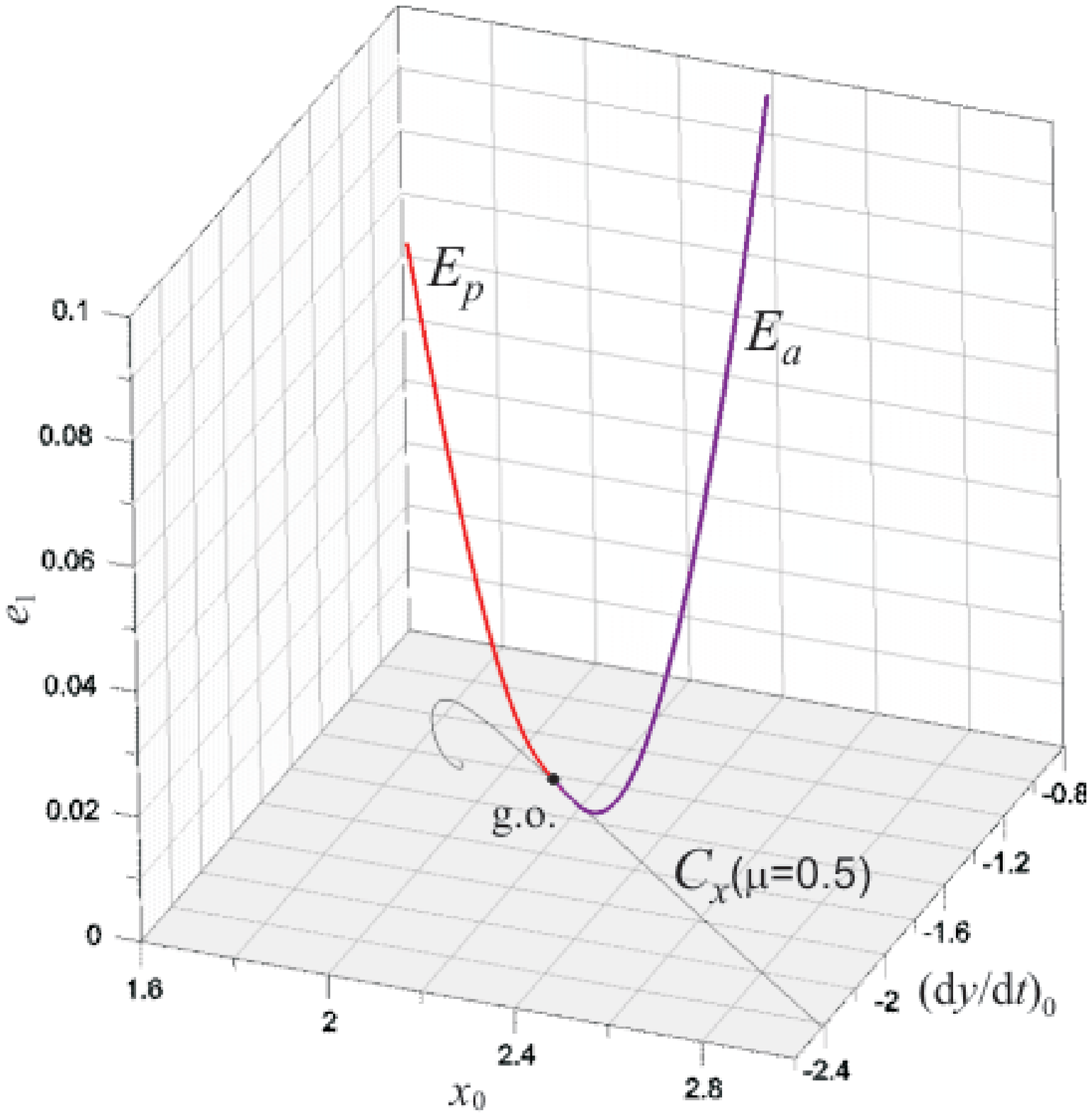} & \includegraphics[width=0.45\hsize]{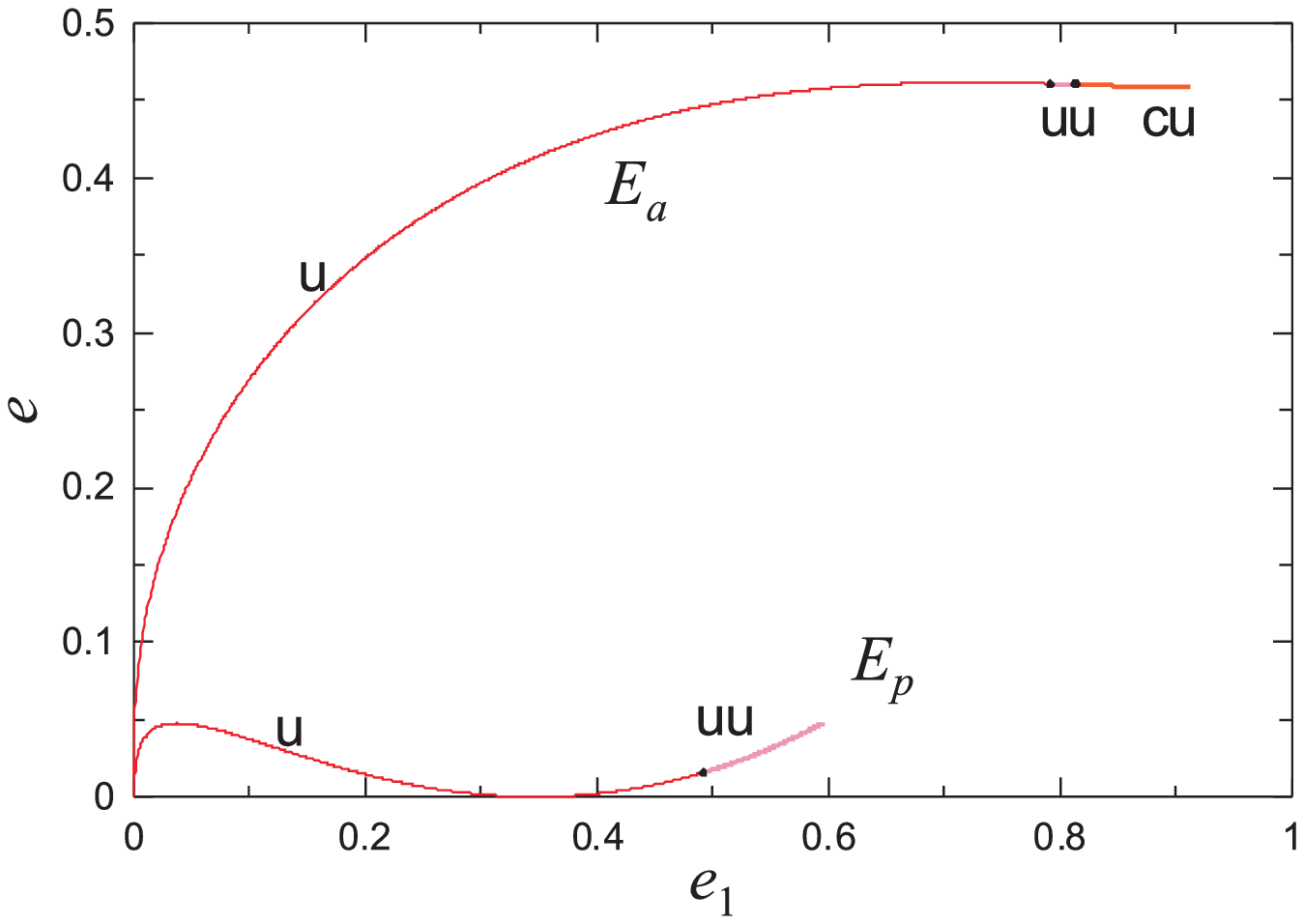}
	\end{array}
	$
	\caption{The $3/2$ resonant families $E_p$ and $E_a$ for $\mu=0.5$. (left) presentation in the space $\Pi'_3=\{(e_1,x_{0},\dot y_{0})\}$. The family of the circular problem ($e_1=0$) and its generating orbit (g.o.) with period $T=6\pi$ is also indicated. (right) the eccentricity, $e$, of the circumbinary orbits of families $E_p$ and $E_a$ and their stability type.} 
	\label{FigEfams}
\end{figure}
\begin{figure}
	\centering
	\includegraphics[width=0.9\hsize]{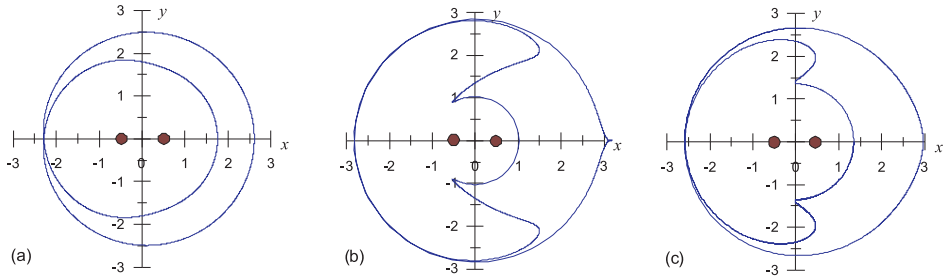}
	\caption{Periodic orbits for $\mu=0.5$ of the families $E_a$, $E_p$ a) orbit of $E_a$ for $e_1=0.1$  b) orbit of $E_a$ for $e_1=0.9$  c) orbit of $E_p$ for $e_1=0.59$. The solid circles indicate the fixed position of the primaries.} 
	\label{FigEorbitsRot}
\end{figure}
\begin{figure}
	\centering
	\includegraphics[width=0.9\hsize]{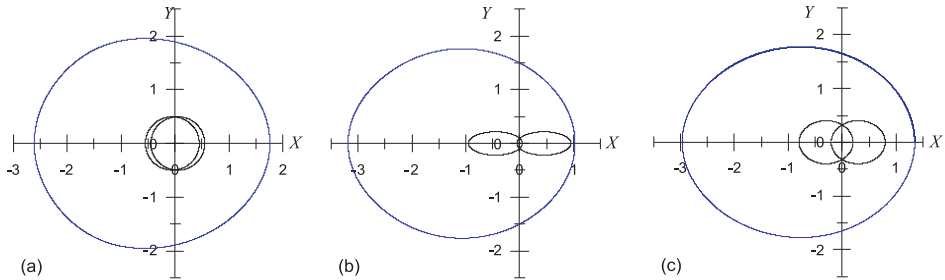}
	\caption{The periodic orbits of Fig. \ref{FigEorbitsRot} in the inertial frame. The orbits of the bodies of the binary are also shown.} 
	\label{FigEorbitsInert}
\end{figure}

In Fig. \ref{FigEorbitsRot} (panels a,b) we show two orbits of family $E_a$ in the rotating frame, where the bodies of the binary are fixed at position $(\pm 0.5,0)$. As $e_1\rightarrow 1$ the orbit of the massless body, $P_2$, tends to collide with the orbit $P_1$ of the binary. However, the continuation of the family breaks at $e_1\approx 0.9$ since the orbit show a cusp at the initial condition $x_0\approx 3.2$ (see panel b)\footnote{In such cases continuation should be computed for initial conditions at a different intersection of the orbit with the $y$ axis}. In panel (c) we show the terminating orbit of the family $E_p$. Although the orbit does not pass close to any primary body, the family breaks due to appearance of strong instability. If we change slightly the initial conditions,  the orbit escapes from the system rapidly. The above orbits are presented also in the inertial frame in Fig. \ref{FigEorbitsInert}. We remark that the periodic orbits of the ERTBP, though they are computed in the rotating frame, they are periodic also in the inertial frame.

\section{Continuation of circular periodic orbits in the GTBP}
For the GTBP of planetary type, where the masses of the bodies $P_1$ and $P_2$ are of the order of Jupiter's mass or less, the circular families  depend in first order approximation on the mass ratio $\rho=m_2/m_1$. The segment of the circular family between the resonances $5:1$ and $2:1$ is presented in Fig. \ref{FigCFamMap}a for some values of the mass ratio $\rho$. All circular orbits are linearly stable except those which belong to a short section of the families in the neigbourhood of the $3:1$ resonance.  Conclusively, as we mentioned in section \ref{SecPOS}, planetary orbits of low eccentricities should be stable. Such a stability domain is depicted e.g. in Fig. \ref{FigCFamMap}b, where a stability map is presented, based on the computation of the maximum Lyapunov characteristic number (LCN). 
\begin{figure}
	\centering
	\includegraphics[width=1.0\hsize]{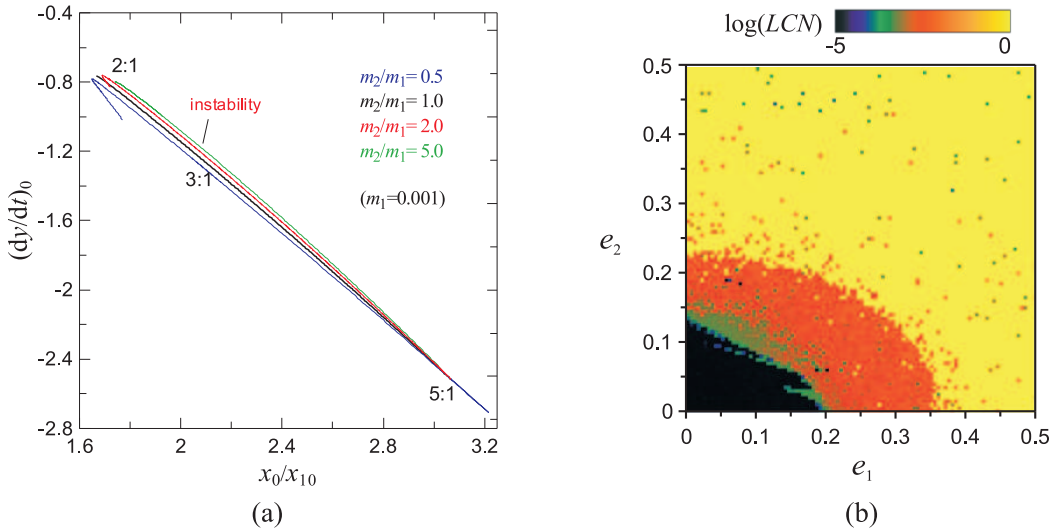}
	\caption{a) Families of circular periodic orbits of the GTBP of planetary type for some planetary mass ratio values b) Stability map on the plane of planetary eccentricities around the circular orbit at $x_0/x_{10}\approx 1.7$ and for $m_1=m_2=0.001$. Dark color represents stable motion.} 
	\label{FigCFamMap}
\end{figure}

Starting from the orbits of the above mentioned families, we can form a binary system of a Jupiter-like planet by performing continuation with respect  to mass, either $m_1$ or $m_2$. Then any orbit of this $\mu-\mathrm{family}$ can be continued in phase space by varying $r_0$ or, equivalently, the initial position $x_{10}$ of $P_1$. 

\subsection{Continuation with respect to planetary mass}
We consider as a starting point of our numerical computations an orbit located at the resonance $5:2$ of the circular family for $m_1=m_2=0.001$. We perform continuation by increasing the mass $m_1$ of the inner planet and keeping $m_2$ fixed. Thus, the bodies $P_0$ and $P_1$ are considered as the primaries of the binary system, while the outer body $P_2$ is a Jupiter-like planet. We obtain that the circular shape of the orbits is preserved along the $m_1-\mathrm{family}$. The initial position ratio $x_0/x_{10}$, the $\dot y_0$ and the energy-Jacobi integral $E$ along this family is presented in Fig. \ref{FigM1Fam}. As $m_1$ increases the radius of the orbit of the primary $P_1$ remains relatively constant but the orbit of $P_2$ is pushed to larger radii and is always a circumbinary orbit (P-type orbit). All these orbits are linearly stable. 
\begin{figure}
	\centering
	\includegraphics[width=1.0\hsize]{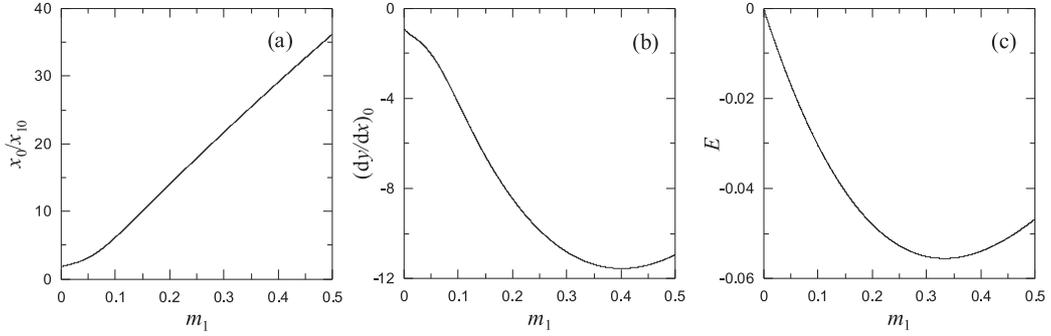}
	\caption{The variation of initial conditions a) $x_0/x_{10}$  b) $\dot y_0$   c) energy $E$ along the $m_1-\mathrm{family}$ (in computations we used $m_2=0.001$, $x_{10}=1.3326$)} 
	\label{FigM1Fam}
\end{figure}
\begin{figure}
	\centering
	\includegraphics[width=1.0\hsize]{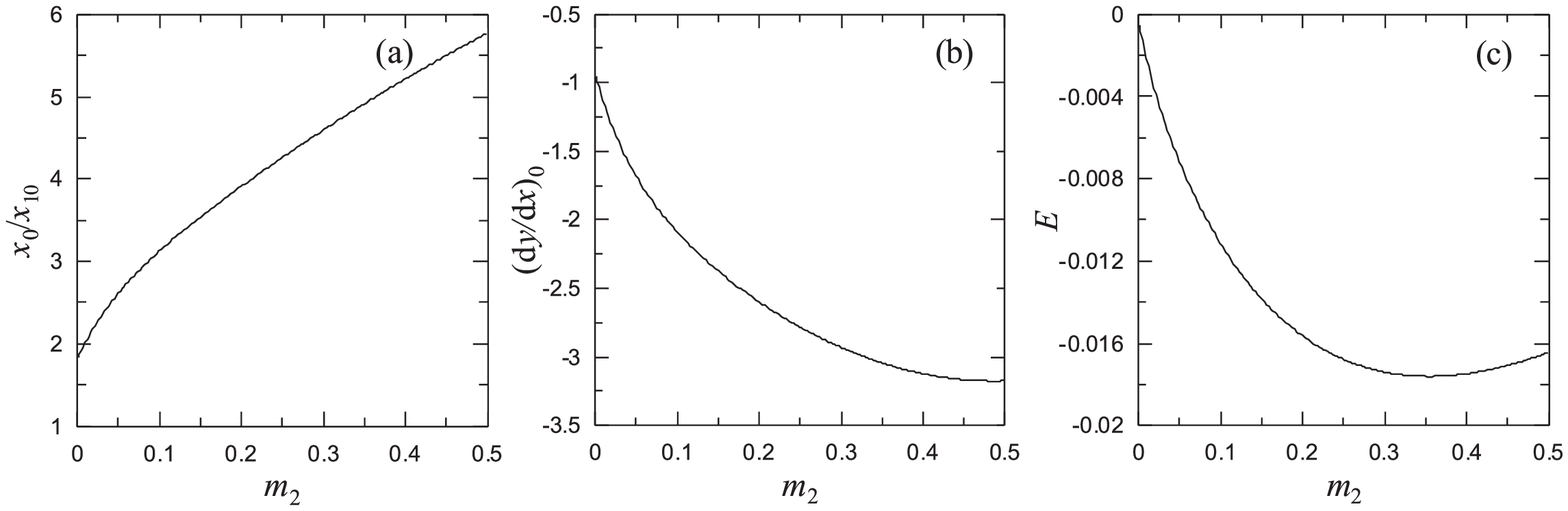}
	\caption{As Fig. \ref{FigM1Fam} for the $m_2-\mathrm{family}$.} 
	\label{FigM2Fam}
\end{figure}

Starting from the same orbit as above, we perform now continuation by increasing the mass $m_2$ of the outer planet $P_2$ (the second primary of the binary) and keeping the mass of the planet $P_1$ fixed. The computed $m_2-\mathrm{family}$ is presented in the plots of Fig. \ref{FigM2Fam}. Some orbits along the families are depicted in the inertial frame in Fig. \ref{FigM2Orbits}. For small mass $m_2$ (of planetary order), the orbits of $P_1$ and $P_2$ are almost circular. However, as the mass of $P_2$ increases, we observe a significant perturbation to the orbit of $P_1$ but the orbits remain linearly stable. Also, the orbit of the planet occupy a ring, which, as $m_2$ increases, becomes a disk around the center of mass. Then gravitational interaction with $P_0$ (the single star at the starting point) becomes strong and the planet $P_1$ turns to become a satellite of $P_0$ (S-type orbit).    
\begin{figure}
	\centering
	\includegraphics[width=1.0\hsize]{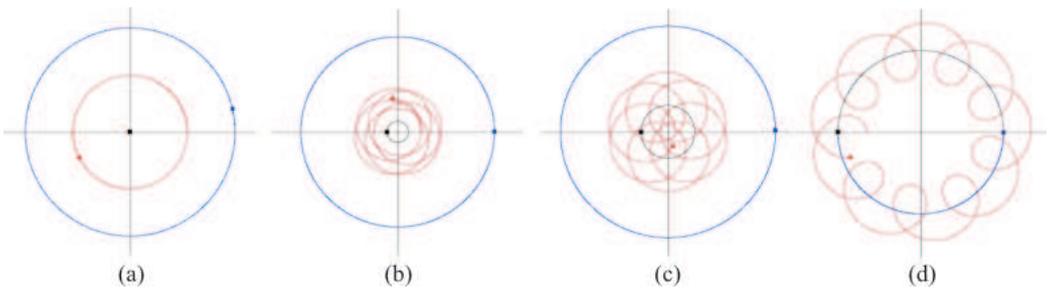}
	\caption{Orbits in the inertial frame of the bodies of the system which correspond to orbits of the $m_2-\mathrm{family}$ at  a) $m_2=0.001$  b) $m_2=0.1$  c) $m_2=0.2$ and d) $m_2=0.5$. For all cases $m_1=0.001$. The orbits of $P_0$, $P_1$ and $P_2$ are shown with black, red, and blue color, respectively. The integration time is equal to the time of one revolution of $P_2$.} 
	\label{FigM2Orbits}
\end{figure}       

\subsection{Continuation with fixed planetary masses}
All orbits of the $m_1$ or $m_2-\mathrm{family}$ computed above, are continued in phase space by varying $x_{10}$. As an example, we take as a starting orbit the orbit of $m_1-family$ at $m_1=0.2$, while $m_2=0.001$. By decreasing $x_{10}$, $x_0$ increases and the orbit of the planet becomes very distant from the binary. Also, it remains circular and linearly stable. Continuation to the other direction (increasing $x_{10}$), the planetary orbit becomes closer to the orbit of $P_1$ and becomes unstable for $x_0/x_{10}<2.6$ (see Fig. \ref{FigM1xcont}).  As the initial position of the planet approaches the binary, the unstable circular orbit deviates significantly from its circular shape in the inertial frame. Finally, the continuation process terminates at a very unstable orbit. This takes place when the planet comes to the $3:1$ resonance with the binary. All orbits are P-type orbits. 
\begin{figure}
	\centering
	\includegraphics[width=1.0\hsize]{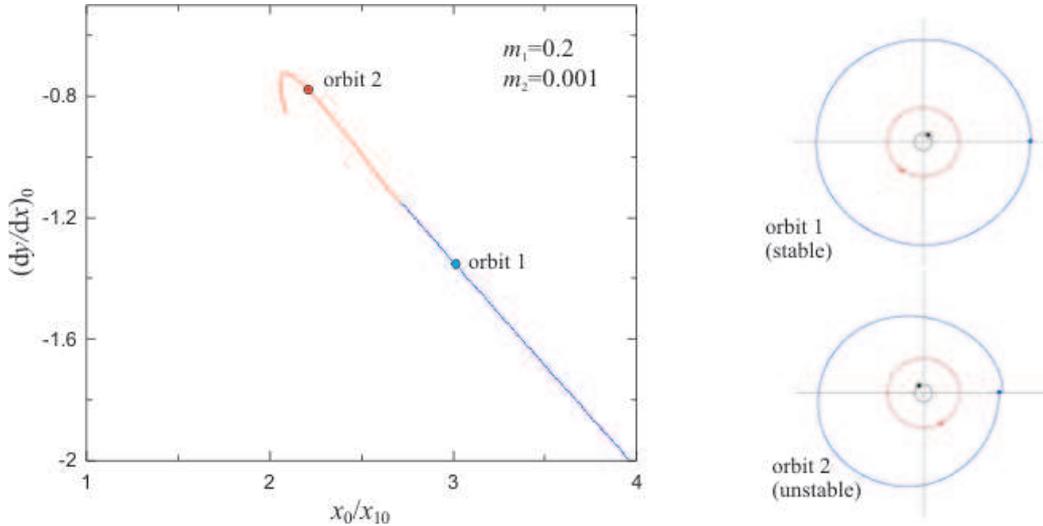}
	\caption{The circular family of periodic orbits for $m_1=0.2$ ($m_2=0.001$). The family continues for $x_0/x_{10}\rightarrow \infty$ but breaks when $x_0/x_{10}\rightarrow 2.09$. Two sample orbits are presented in the inertial frame on the right panels.} 
	\label{FigM1xcont}
\end{figure}   

Starting from orbits of the $m_2-\mathrm{family}$, the planet is the body $P_1$ ($m_1=0.001$). In Fig. \ref{FigM2xcont} we show the family, which is constructed after $x$-continuation  starting from the orbit at $m_2=0.25$. All orbits are S-type orbits, namely $P_1$ is a satellite of the heavy body $P_0$. The orbits are linearly stable along the family (at least for the presented segment). Also, two sample orbits (at the left and right side of the family) are presented in the inertial frame. As $x_{10}$ decreases, the orbit of $P_1$ comes closer to $P_0$ and revolves with high frequency with respect to the rotation frequency of the binary $P_0P_2$. This is concluded also by the angular velocity $\dot \theta$ of the rotating frame shown in panel (c).        
 \begin{figure}
	\centering
	\includegraphics[width=1.0\hsize]{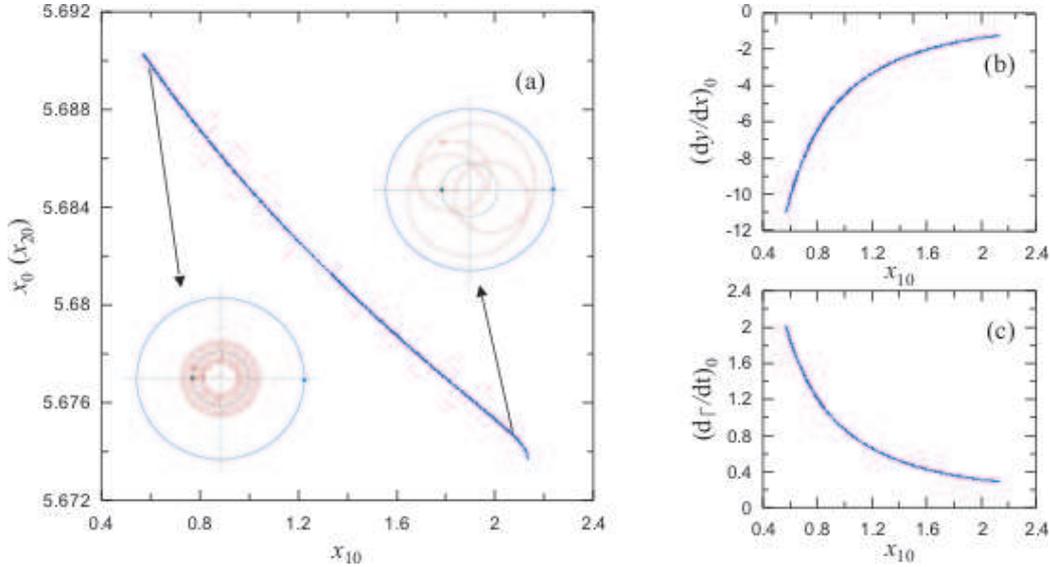}
	\caption{The circular family of periodic orbits for $m_2=0.25$ ($m_1=0.001$). Presentation in the planes a) $x_{10}-x_{20}$, where $x_{20}\equiv x_0$ is the position of the second primary body $P_2$  and b) $x_{10}-\dot y_{0}$.  c) The angular frequency $\dot\theta_0=\dot\theta(0)$ of the rotating frame along the family.} 
	\label{FigM2xcont}
\end{figure}          
 
\section{Conclusions}
In this paper we review theoretical and computational aspects for the computation and continuation of periodic orbits in the framework of the planar three body problem. We show that continuation with respect to the mass can be used for known solutions of unperturbed or planetary configurations. When the mass of a small body increases to large values of the order of the original primary, we obtain a binary system and a planetary periodic orbit of $P$ or $S$-type. These periodic orbits can be continued in phase space keeping fixed the masses of the two primaries and families of periodic solutions are constructed. Their linear stability can be determined from the monodromy matrix. 

Our computations have been limited only to families of circular periodic solutions. Starting from the unperturbed system we showed how to compute circumbinary periodic solutions in the CRTBP. The family of such solutions exists for planetary orbits of radius up to infinity and they are linearly stable. But as the  radius of the planetary orbit becomes smaller and smaller along the family, the gravitational interaction of the planet with the outer primary becomes significant, the shape of the orbit deviates from its circular geometrical form and becomes unstable. The family terminates when the planetary orbit approaches the orbit of the outer primary and the motion becomes strongly unstable. 

Continuation of circular periodic orbits of the GTBP are also presented. We obtained stable orbits for a planet of the mass of Jupiter, $m_i=0.001$, in a binary system of total mass $m_0+m_j=0.999$. We started from a two planet system and performed continuation with respect to the mass of the inner or the outer planet. When we increase the mass of the inner planet circumbinary orbits are obtained. Instead, when the mass of the outer planet is increased, circumstellar orbits are formed.

The methodology described can be applied also for elliptic binary systems and linear stability can be compared with the stability limits obtained by \citet{Dvorak02} and \citet{Musielak05}. Also, triple systems, where all bodies are of masses of the same order, can be approached by the method of {\em mass continuation} and stability results can be extracted for such systems.  
 
\vskip 0.5cm
\noindent
{\bf acknowledgements} This work is dedicated to the memory of J.D. Hadjidemetriou who was a pioneer in the computation of periodic orbits in the general three body problem.

\bibliographystyle{aasjournal}
\bibliography{gv00nbib}

\end{document}